\documentclass[11pt]{article} 
\usepackage{chicagor} 

\usepackage{latexsym} 
\usepackage{amsmath} 
\usepackage{amssymb} 

\allowdisplaybreaks[2]

\ifx\pdfoutput\undefined 
\else 
\fi

\newcommand{\sat}{\models} 
\newcommand{\rimp}{\Rightarrow} 
 
\renewcommand{\phi}{\varphi}

\renewcommand{\emptyset}{\varnothing} 
 
\newcommand{\intension}[1]{[\![ #1 ]\!]} 
\newcommand{\event}[2]{\intension{#1}_{#2}}

\newcommand{\true}{\mathbf{true}} 
\newcommand{\false}{\mathbf{false}} 

\newcommand{\truep}{\mathit{true}} 
\newcommand{\falsep}{\mathit{false}}

 \newcommand{\AX}{\text{AX}} 
\newcommand{\A}{\mathtt{A}} 
\newcommand{\cA}{\mathcal{A}} 
\newcommand{\cC}{\mathcal{C}}

\newcommand{\Ver}{\mi{Ver}} 
\newcommand{\Prob}{\mi{Prob}} 
\newcommand{\WVer}{\mi{WVer}} 
\newcommand{\DC}{\mi{DC}} 
\newcommand{\KC}{\mi{KC}} 
\newcommand{\KL}{\mi{KL}} 
\newcommand{\imp}{\mi{imp}}

\usepackage{amsthm} 

\newtheoremstyle{theorem}{\topsep}{\topsep}%
     {\itshape}%
     {}%
     {\bfseries}%
     {:}%
     {10pt}%
     {\thmname{#1}\thmnumber{ #2}\thmnote{ (#3)}}%

\theoremstyle{theorem} 
\newtheorem{theorem}{Theorem}[section] 
 
\newtheorem{corollary}[theorem]{Corollary} 
\newtheorem{proposition}[theorem]{Proposition}

\newtheorem{EXAMPLE}[theorem]{Example} 
\newenvironment{example} 
   {\begin{EXAMPLE}\rm} 
   {\mbox{}\qedsymbol\end{EXAMPLE}} 

\newcommand{\bbox}{\vrule height7pt width4pt depth1pt} 
\renewcommand{\qedsymbol}{\bbox}

\newcommand{\Prime}{\mathit{Prime}}

\newcommand{\derivesDY}{\vdash_{\scriptscriptstyle DY}} 
\newcommand{\msg}{m} 
\newcommand{\ky}{k} 
\newcommand{\encr}[2]{\{#1\}_{#2}} 
\newcommand{\Rule}[2]{          %
  \begin{array}{c} 
  #1 \\\hline 
  #2 
  \end{array}} 
\newcommand{\mi}[1]{\mathit{#1}} 
\newcommand{\ADY}{\A^{\scriptscriptstyle\rm DY}} 
\newcommand{\commentout}[1]{} 
\newcommand{\dimp}{\Leftrightarrow} 
\newcommand{\LK}{\mathcal{L}^K} 
\newcommand{\LKQU}{\mathcal{L}^{K,\mi{QU}}} 
\newcommand{\PlaySports}{\text{$\mi{Play}$-$\mi{sports}$}} 
\newcommand{\Sub}{\mi{Sub}}

\newcommand{\subbit}[1]{\subsection{#1}}

\newenvironment{namedaxiom}[1]
  {\renewcommand*{\theequation}{\textit{#1}}%
   \addtocounter{equation}{-1}%
   \begin{equation}}
  {\end{equation}} 

\begin{document} 

\title{Dealing With Logical Omniscience:\\ Expressiveness and Pragmatics} 

\author{Joseph Y. Halpern\\ 
Cornell University\\ 
Ithaca, NY 14853 USA\\ 
halpern@cs.cornell.edu 
\and
Riccardo Pucella\\ 
Northeastern University\\ 
Boston, MA 02115 USA\\ 
riccardo@ccs.neu.edu} 

\date{}

\maketitle 

\begin{abstract} 
We examine four approaches for dealing with the logical 
omniscience problem and their potential applicability: 
the syntactic approach, awareness, algorithmic knowledge, and 
impossible possible worlds. Although in some settings these approaches are 
equi-expressive and can capture all epistemic states, in other 
settings of interest (especially with probability in the picture), we 
show that they are not equi-expressive.   
We then consider  
the 
pragmatics of dealing with logical omniscience---%
how to choose an approach and construct an appropriate model. 
\end{abstract} 

\section{Introduction}

Logics of knowledge based on possible-world semantics are useful in 
many areas of knowledge representation and reasoning, 
ranging from security to distributed computing to game theory. 
In these models, an agent is said to know a fact $\phi$ if $\phi$ is 
true in all the worlds she considers possible. 
While reasoning about knowledge with this semantics has proved 
useful, 
as is well known, it suffers from  
what is 
known in the literature as the \emph{logical omniscience} problem: 
under possible-world semantics, agents know all tautologies and 
know the   logical consequences of their knowledge.   

While logical omniscience is certainly not always an issue,  
in many applications it is.  For example,  
in the context of distributed computing, we are interested in 
polynomial-time algorithms, although in some cases the knowledge 
needed to perform optimally may require calculations that cannot be 
performed in polynomial time (unless P=NP) \cite{MT};  
in the context of security, we may want to reason about 
computationally bounded adversaries who cannot factor a 
large composite number, and thus cannot be logically omniscient;  
in game theory, we may be interested in the impact of computational 
resources on solution concepts (for example, what will agents do if 
computing a Nash equilibrium is difficult). 

Not surprisingly, many approaches for 
dealing with the logical omniscience problem  have been suggested  
(see \cite[Chapter 9]{r:fagin95} and \cite{r:moreno98}).  A far 
from exhaustive  
list of  approaches includes: 
\begin{itemize} 
\item syntactic approaches \cite{Eb,MoHe,Kon1}, where an agent's knowledge is 
represented by a set of formulas (intuitively, the set of formulas she 
knows);  
\item \emph{awareness} \cite{FH}, where 
an agent knows $\phi$ if she is aware of $\phi$ and $\phi$ is true in 
all the worlds she considers possible;   
\item \emph{algorithmic knowledge} 
\cite{r:halpern94} 
where, roughly 
speaking, an  
agent knows $\phi$   
if her knowledge algorithm returns $\text{``Yes''}$ on a query of 
$\phi$; and 
\item \emph{impossible worlds} \cite{Rant}, where  
the agent may  consider possible worlds that are logically 
inconsistent (for  
example, where $p$ and $\neg p$ may both be true). 
\end{itemize} 

Which approach is best to use, of course, depends on the 
application.  
One goal of this paper is to elucidate the aspects   
of the application that make a logic more or less appropriate.   
We start by considering the expressive power of these approaches. 
It may seem that there is not much to say with regard to expressiveness, 
since it has been shown that all these approaches are equi-expressive 
and, indeed, can capture all epistemic states 
(see \cite{Wansing,r:fagin95} and Section~\ref{sec:review}). 
However, this result holds only if we allow an agent to consider no 
worlds possible. 
As we show, this equivalence no longer holds in contexts where 
agents must consider some worlds possible.  This is particularly  
relevant once we have  
probability in the picture. 
But expressive power is only part of the story. 
We consider here (mainly by 
example) the \emph{pragmatics} of dealing with logical omniscience---an issue 
that has largely been ignored: 
how to choose an approach and construct an appropriate model. 
\section{The Four Approaches: A Review}\label{sec:review}  

We now review the standard possible-worlds approach and the four  
approaches to dealing  logical omniscience discussed in the  
introduction. 
For ease of exposition we focus on the single-agent  
propositional 
case. 
While in many applications it is important to consider more than one 
agent 
and to allow first-order features (indeed, this is true in some of our 
examples),  
the issues that arise in dealing with multiple agents 
and first-order features
are largely 
orthogonal to those involved in dealing with logical omniscience.   
Thus, we do not discuss these extensions here.
\newcommand{\Ms}{M}     %
\newcommand{\Maw}{M}   %
\newcommand{\Mal}{M}    %
\newcommand{\Mi}{M}     %

\subbit{The Standard Approach} 
Starting with a set $\Phi$ of propositional formulas, we close off under 
conjunction, negation, and the $K$ operator.   
Call the resulting language $\LK$. 
We give semantics to these 
formulas using Kripke structures.   
For simplicity, we focus on approaches that satisfy the K45 axioms (as 
well as KD45 and S5). 
In this case, a  
\emph{K45 Kripke structure} is a triple $(W,W',\pi)$, where $W$ is a 
nonempty set of \emph{possible worlds} (or \emph{worlds}, for short),  
$W'\subseteq W$ is the set of worlds that the agent considers possible, and 
$\pi$ is an \emph{interpretation} that associates with each world a 
truth assignment  
$\pi(w)$ to the primitive propositions in $\Phi$.   
Note that the agent need not consider every possible world  
(that is, each world in $W$) 
possible.  
Then we have 
\begin{itemize} 
\item[] $(M,w) \sat p$ iff $\pi(w)(p) = \true$ if $p \in \Phi$. 
\item[] $(M,w) \sat\neg\phi$ iff $(M,w)\not\sat\phi$. 
\item[] $(M,w) \sat\phi\land\psi$ iff $(M,w)\sat\phi$ and $(M,w)\sat\psi$. 
\item[] $(M,w)\sat K\phi$ iff $(M,w')\sat\phi$ for all $w'\in W'$. 
\end{itemize} 

This semantics suffers from the logical omniscience problem.  In 
particular, one sound axiom is 
\[(K\phi \land K(\phi \rimp \psi)) \rimp K\psi,\]
which says that an agent's knowledge is closed under implication. 
In addition, the \emph{knowledge generalization} inference rule is 
sound: 
\[ \text{From  $\phi$ infer  $K \phi$.}\]
Thus, 
agents know all tautologies. 
As is well known, two other axioms are sound  
in K45 Kripke structures: 
\[ K\phi \rimp KK\phi \]
and 
\[ \neg K\phi \rimp K\neg K\phi.\]
These are  
known respectively as the positive and negative introspection axioms. 
(These properties  
characterize 
K45.)  

In the structures we consider, we allow $W'$ to be empty, in which case 
the agent  
does not consider any worlds possible. 
In such structures, the formula $K(\falsep)$ is true. 
A \emph{KD45 Kripke structure} is a K45 Kripke structure $(W,W',\pi)$ 
where  
$W'\ne\emptyset$. Thus, in a KD45 Kripke structure, the agent always 
considers at least one world possible.  In KD45 Kripke structures,  
the axiom 
\[ \neg K(\falsep)\]
is sound,  
which implies that the agent cannot know inconsistent facts.  
The logic KD45 results when we add this axiom to K45. 
\emph{S5 Kripke structures} are KD45 Kripke structures where $W = W'$; 
that is, the agent considers all worlds in $W$ possible. 
In S5 Kripke structures, 
the axiom  
\[ K \phi \rimp \phi,\]
which says that the agent can know only 
true facts, is sound. 
Adding this axiom to the KD45 axioms gives us the logic S5.

\subbit{The Syntactic Approach} 
The intuition behind the syntactic approach for dealing with logical 
omniscience is simply to 
explicitly list, at every possible  
world $w$, the set of formulas that the  agent knows at $w$. 
A \emph{syntactic structure} has the form 
$\Ms=(W,W',\pi,\cC)$, where $(W,W',\pi)$ is a  
K45 
Kripke structure and 
$\cC$ associates a set  of formulas $\cC(w)$ with every world $w\in W$.   
The semantics of primitive propositions, conjunction, and negation is 
just the same as for Kripke structures.  For knowledge, we have 
\begin{itemize} 
\item[]  
$(\Ms,w)\sat K\phi$ iff $\phi\in\cC(w)$. 
\end{itemize}

\subbit{Awareness} 
Awareness is based on the intuition that an agent should be aware of a  
concept before she can know it.  The formulas that 
an agent is aware of are represented syntactically; 
we associate with every world 
$w$ the set $\cA(w)$ of formulas that the agent is aware of.  
For an agent to know a formula $\phi$, not only does $\phi$  
have to be true at all the worlds she considers possible, but she has 
to be aware of $\phi$ as well.  
A \emph{K45 awareness structure} is a tuple  
$\Maw = (W,W',\pi,\cA)$, where  
$(W,W',\pi)$ is a K45 Kripke structure and $\cA$ maps worlds to  
sets of formulas.   
We now define 
\begin{itemize} 
\item[] $(\Maw,w)\sat K\phi$ iff $(\Maw,w')\sat \phi$ for all $w' \in W'$ 
and $\phi\in\cA(w)$.%
\footnote{In \cite{FH}, the symbol $K$ is reserved for the standard 
definition of knowledge; the definition we have just given is denoted as 
$X\phi$, where $X$ stands for \emph{explicit} knowledge.   
A similar remark applies to the algorithmic knowledge approach below.  
We use $K$ throughout for ease of exposition.} 
\end{itemize} 
We can define KD45 and S5 awareness structures in the obvious way: 
$M=(W,W',\pi,\cA)$ is a KD45 awareness structure when 
$(W,W',\pi)$ is a KD45 structure, and an S5 awareness structure when 
$(W,W',\pi)$ is an S5 structure. 

\subbit{Algorithmic Knowledge} 
In some applications, there is a computational intuition underlying 
what an agent knows; that is, an agent computes what she knows using an 
algorithm.   
\emph{Algorithmic knowledge} is one way of formalizing this intuition. 
An \emph{algorithmic knowledge structure} is a tuple $\Mal = 
(W,W',\pi, \A)$, where $(W,W',\pi)$ is a  
K45 
Kripke structure and $\A$ 
is a \emph{knowledge algorithm} that returns 
$\text{``Yes''}$, $\text{``No''}$, or $\text{``?''}$  
given a formula $\phi$.%
\footnote{In \cite{r:halpern94}, the knowledge algorithm is also given 
an argument that describes the agent's local state, which, roughly 
speaking, captures the relevant information that the agent has. 
However, in our single-agent static setting, there is only one 
local state, so this argument is unneeded.} 
Intuitively, $\A(\phi)$ returns ``Yes'' if the agent can compute that 
$\phi$ is true, ``No'' if the agent can compute that $\phi$ is false, 
and ``?'' otherwise. 
In algorithmic  
knowledge 
structures, 
\begin{itemize} 
\item[] $(\Mal,w)\sat K\phi$ iff $\A(\phi)=\text{``Yes''}$. 
\end{itemize} 

An important class of knowledge algorithms consists of the 
\emph{sound} knowledge algorithms.  
When a sound knowledge algorithm returns $\text{``Yes''}$ to a query 
$\phi$, then the agent knows (in the standard sense) $\phi$, and when 
it returns $\text{``No''}$ to a query $\phi$, then the agent does 
not know (again, in the standard sense) $\phi$.  
Thus, if $\A$ is a sound knowledge algorithm, then 
$\A(\phi)=\text{``Yes''}$ implies $(M,w)\sat\phi$ for all $w\in W'$, 
and and $\A(\phi)=\text{``No''}$ implies there exists $w\in W'$ such 
that $(M,w)\sat\neg\phi$.  
(When $\A(\phi)=\text{``?''}$, nothing is prescribed.)

\subbit{Impossible Worlds} 
The impossible-worlds approach relies on relaxing the notion of 
possible world.  Take the special case of logical omniscience that says 
that an agent knows all tautologies.  
This is a consequence of the fact that a tautology must be true at 
every possible world. 
Thus, one way to eliminate this problem is to allow tautologies to be 
false at some worlds.  
Clearly, those worlds do not obey the usual laws of logic---they are 
\emph{impossible possible worlds} 
(or \emph{impossible worlds}, for short).  

A \emph{K45} (resp., \emph{KD45}, \emph{S5}) \emph{impossible-worlds structure} is a tuple 
$\Mi=(W,W',\pi,\cC)$, where $(W,W'\cap W,\pi)$ is a  
K45 (resp., KD45, S5) Kripke 
structure, $W'$ is the set of worlds that the agent considers
possible, and $\cC$ associates with each world in $W'-W$ a set of
formulas.   
$W'$, the set of worlds the agent considers possible, is not 
required to be a subset of $W$---the agent may well include 
impossible worlds in $W'$.  
The worlds in $W'-W$ are the impossible worlds.  
We can also consider a class of impossible-worlds structures 
intermediate between K45 and KD45 impossible-worlds structures. 
A \emph{KD45$^-$ impossible-worlds structure} is a K45 impossible-worlds 
structure $(W,W',\pi,\cC)$ where $W'$ is nonempty.  In a 
KD45$^-$  
impossible-worlds  
structure, we do not require that $W' \cap W$ be nonempty. 
A formula $\phi$ is true at a world $w \in W'-W$ if and only if $\phi 
\in \cC(w)$; for worlds $w \in W$, the truth assignment is like that in 
Kripke structures.  
Thus, 
\begin{itemize} 
\item if $w \in W$, then $(\Mi,w) \sat p$ iff $\pi(w)(p) = \true$; 
\item if $w \in W$, then $(\Mi,w) \sat K_i \phi$ iff $(\Mi,w') \sat \phi$ for 
all $w' \in W'$; 
\item if $w \in W'-W$, then $(\Mi,w)\sat\phi$ iff $\phi \in \cC(w)$. 
\end{itemize} 
We remark that when we speak of validity in impossible-worlds 
structures, we mean truth at all  
possible worlds in $W$ 
in all impossible-worlds structures $M = (W, \ldots)$.

\section{Expressive Power}\label{sec:expressive} 

There is a sense in which all four approaches are equi-expressive, and 
can capture all states of knowledge. 
\begin{theorem}\label{thm:equivalence}  
{\rm \cite{Wansing,r:fagin95}} For 
every finite set $F$ of formulas
and every propositionally consistent set $G$ of formulas, 
there exists a syntactic structure (resp., K45 awareness structure,  
KD45$^-$ impossible-worlds structure, algorithmic knowledge structure) 
$M = (W, \ldots)$ and a world $w \in W$ such that $(M,w) \sat K \phi$ 
if and only if $\phi \in F$,
and $(M,w)\sat\psi$ for all $\psi\in G$.\footnote{This result extends
to infinite sets $F$ of formulas for syntactic structure, K45
awareness structures, and KD45$^{-}$ impossible-worlds structures. For 
algorithmic knowledge structures,  the result extends to recursive
sets $F$ of formulas.} 
\end{theorem} 
\begin{proof} We review the basic idea of the proof, since it will set
the stage for our later results.  
\begin{itemize} 
\item For syntactic structures, let $M=(\{w\},\emptyset,\pi,\cC)$, where
$\cC(w) = F$ and $\pi(w)$ is such that $(M,w) \sat \psi$ for all $\psi \in
G$.  (Since $G$ is propositionally consistent, there must be a truth
assignment that makes all the formulas in $G$ true; we can take $\pi(w)$
to be that truth assignment.)
\item For K45 awareness structure, let $M=(\{w\},\emptyset,\pi,\cA)$, where
$\cA(w)=F$ and $\pi(w)$ makes all the formulas in $G$ true.
\item For KD45$^-$ impossible-worlds structure, let
$M=(\{w\},\{w'\},\pi,\cC)$, where $\cC(w') = F$
and $\pi(w)$ makes all the formulas in $G$ true. 
\item For algorithmic knowledge, let $M=(\{w\},\emptyset,\pi,\A)$,
where $\A(\phi) = \text{``Yes''}$ iff $\phi \in F$ and
$\pi(w)$ makes all the formulas in $G$ true. 
\end{itemize} 
 \end{proof}

Despite the name, the introspective axioms of K45 are not 
valid in K45 awareness structures or K45 impossible-worlds structures. 
Indeed, it follows from Theorem~\ref{thm:equivalence} that no axioms of 
knowledge are valid in these structures.   
(Take $F$ to be the empty set.)   
To make this precise, 
let $\mi{Prop}$ be the axiom 
\begin{namedaxiom}{Prop}
\text{$\phi$ is a valid formula of propositional 
logic}
\end{namedaxiom}
and $\mi{MP}$ be the inference rule
\begin{namedaxiom}{MP}
\text{From $\phi\rimp\psi$ and $\phi$ infer $\psi$.}
\end{namedaxiom}

\begin{theorem}\label{thm:axioms} 
$\{\mi{Prop},\mi{MP}\}$ is a sound and complete 
axiomatization of $\LK$ with respect to K45 
awareness structures (resp., K45 and KD45$^{-}$ 
impossible-worlds structures, syntactic structures, algorithmic 
knowledge structures). 
\end{theorem} 

\begin{proof} 
Suppose that $\phi$ is consistent with $\{\mi{Prop},\mi{MP}\}$.  It suffices
to show that $\phi$ is satisfiable in a K45 awareness (resp., K45 and
KD45$^{-}$ impossible-worlds structure, syntactic structure, algorithmic 
knowledge structure).  Viewing formulas of the form $K\psi$ as primitive
propositions, $\phi$ must be propositionally consistent.  Thus, there
must be a truth assignment $v$ to the primitive propositions and formulas of
the form $K\psi$ that appear in $\phi$ such that $\phi$ evaluates to
true under this truth assignment.  Let $F$ consist of all formulas 
$\psi$ such that $v(K \psi) = \true$ and let $G$ consist of all the
propositional formulas $\psi$ such that $v(\psi) = \true$. 
Let $M$ be the structure guaranteed to exist by Theorem~\ref{thm:equivalence}. 
It is easy to see that $(M,w) \sat \phi$.
\end{proof} 

It follows from Theorem~\ref{thm:axioms} that a formula is valid with
respect to K45 awareness structures (resp., K45 and KD45$^-$
impossible-worlds structures, syntactic structures, algorithmic
knowledge structures) if and only if it is propositionally valid, if
we treat formulas of the form $K \phi$ as primitive propositions.
Thus, deciding if a formula is valid is co-NP complete, just as it is
for propositional logic.  

Theorems~\ref{thm:equivalence} and~\ref{thm:axioms} rely on the fact 
that we are considering  
K45 awareness structures and KD45$^-$ (or K45) impossible-worlds 
structures.  
(Whether we consider K45, KD45, or S5 is irrelevant in the case of 
syntactic structures and algorithmic knowledge structures, since the
truth of a  
formula does not depend on what worlds an agent considers possible.) 
There are constraints on what can be known if we consider
KD45 and S5 awareness structures and impossible-worlds 
structures.  The constraints depend on which structures we consider.  
To make the constraints precise, we need a few definitions.
We say a set of formulas $F$ is \emph{downward closed} if 
the following conditions hold:
\begin{itemize}
\item[(a)] if $\phi \land \psi \in F$,
then both $\phi$ and $\psi$ are in $F$; 
\item[(b)] if $\neg \neg \phi \in
F$, then $\phi \in F$;
\item[(c)] if $\neg (\phi \land \psi) \in F$, then
either $\neg \phi \in F$ or $\neg \psi \in F$ (or both); and 
\item[(d)]
if $K \phi \in F$, then $\phi \in F$.  
\end{itemize}
We say that $F$ is \emph{k-compatible} with $F'$ if $K \psi \in F'$ implies
that $\psi \in F$.  

\begin{proposition}\label{lem:closure}
Suppose that $M = (W, W', \ldots)$ is a KD45 awareness structure (resp.,
KD45 impossible-worlds structure), $w \in W$, and 
$w' \in W'$ (resp., $w' \in W\cap W'$).
Let $F = \{\phi \mid (M,w) \sat K \phi \}$ and let $F' = \{\psi \mid
(M,w') \sat \psi\}$.  Then
\begin{itemize} 
\item[(a)] $F'$ is propositionally consistent downward-closed set of
formulas that contains $F$;
\item[(b)] if $M$ is a KD45 impossible-worlds structure then
$F$ is k-compatible with $F'$. 
\end{itemize}
\end{proposition}

\begin{proof}
Suppose that $M = (W,W',\ldots)$ is a KD45 awareness structure.
Let $w$, $w'$, $F$, and $F'$ be as in the statement of the theorem.
Clearly $F \subseteq F'$.  Since
$w'$ is a possible world, it is easy to see that $F'$ satisfies the
first three conditions of being downward closed.  For the last
condition, note 
that 
if $(M,w') \sat K \psi$, then we must have $(M,w'') \sat
\psi$ for all worlds $w'' \in W'$, so $(M,w') \sat \psi$.  Finally, 
$F'$ must be propositionally consistent, since $w$ is a possible world.
The argument is the same if $M$ is a KD45 impossible-worlds structure,
since $w' \in W \cap W'$ in this case.  
To see that $F'$ is k-compatible if $M$ is 
a
KD45 impossible-worlds
structure, suppose that $K \phi \in F'$.  By the definition of $F'$,
this means that $(M,w') \sat K\phi$.  It follows that 
$(M,w'') \sat \phi$ for all $\phi \in W'$.  Hence, $(M,w) \sat
K \phi$, so $\phi \in F$.  Note that this argument does not work for
awareness structures, since we may not have $\phi \in \cA(w)$. 
\end{proof}

The next result show that the constraints on $F$ described in
Proposition~\ref{lem:closure} are the only constraints on $F$.

\begin{theorem}\label{t:awareness}    
If $F$ and $F'$ are such that 
$F'$ is propositionally consistent downward-closed set of
formulas that contains $F$, then there exists a KD45 awareness
structure $M = (\{w,w'\}, \{w'\}, \pi, \cA)$ such that $(M,w) \sat K \phi$
iff $\phi \in F$ and $(M, w') \sat \psi$ for all $\psi \in F'$.  If, in
addition, $F$ is k-compatible with $F'$, then there exists a KD45
impossible-worlds structure $M = (\{w,w'\}, \{w',w''\}, \pi, \cC)$ such
that $(M,w) \sat K \phi$ iff $\phi \in F$ and $(M, w') \sat \psi$ for
all $\psi \in F'$.  Finally, if $F = F'$, then we can take $w = w'$, so
that $M$ is an S5 awareness (resp., S5 impossible-worlds) structure.
\end{theorem}
\begin{proof}
In the case of KD45 awareness structures, let $M = (\{w,w'\}, 
\{w'\},
\pi, \cA)$, where  $\pi(w')$ makes all the propositional
formulas in $F'$ true, 
$\cA(w) = F$, and $\cA(w') = \{\phi\mid K \phi \in F'\}$.  We now prove by
induction that if $\phi \in F'$ then $(M,w') \sat \phi$.  This is true by
construction in the case of primitive propositions and follows easily from
the induction hypothesis in the case of conjunctions.  If $\phi$ has the
form $K\psi$ then, since $\psi$ must be in $F'$, it follows from the
induction hypothesis that $(M,w') \sat \psi$ and, by construction, that $\psi
\in \cA(w')$.  Thus, $(M,w') \sat K \psi$.  Finally, if $\phi$ has the
form $\neg \psi$, we consider the possible forms of $\psi$.  If $\psi$
is a primitive proposition it follows from the definition of $\pi(w')$.
If $\psi$ has the form $\neg \psi'$, then $\psi' \in F'$, so, by the
induction hypothesis, $(M,w') \sat \psi'$.  Hence, $(M,w') \sat \phi$.
Similarly, the result follows from the definition of downward closure
and the induction hypothesis if $\psi$ has the form $\psi_1 \land
\psi_2$.  Finally, if $\psi$ has the form $K \psi'$, then the result
follows from the definition on $\cA(w')$.  It is now immediate that 
$(M,w) \sat K\phi$ iff $\phi \in F$: if $(M,w) \sat K \phi$ then it
follows from the definition of $\cA(w)$ that we must have $\phi \in F$.
Conversely, if $\phi \in F$, then $\phi \in \cA(w)$ and $(M,w') \sat
\phi$ (since $F \subseteq F'$), so $(M,w) \sat K \phi$.  

If $F=F'$, then we can take $w = w'$ in this argument
to get an S5 awareness structure.

In the case of impossible-worlds structures, let $M =
(\{w,w'\},\{w',w''\},\pi, \cC\}$, where  $\pi(w')$ makes all the  
propositional formulas in $F'$ true
and $\cC(w'') = F$.  A proof by induction on the
structure of formulas much like that above shows that $(M,w') \sat \phi$
if $\phi \in F'$.  To deal with the case that $\phi = K \psi$, we use
the fact that $F$ is k-compatible
with 
$F'$ to get that $\psi \in F$, so that
$(M,w'') \sat \psi$.  To see that $(M,w) \sat K \phi$ iff $\phi \in F$,
first observe that if $\phi \in F$ then, by construction $\phi \in
\cC(w'')$, and, since $F \subseteq F'$, $(M,w') \sat \phi$, so $(M,w)
\sat K \phi$. For the converse, if $(M,w) \sat K \phi$, then $(M,w'')
\sat \phi$, so $\phi \in F$.  
\end{proof}

We can characterize these properties axiomatically.  Let 
$\mi(Ver)$ (for \emph{Veridicality}) be the standard axiom that
says that everything known must be true: 
\begin{namedaxiom}{Ver}
K \phi \rimp \phi.
\end{namedaxiom}
Let $\AX_{\Ver}$ be the axiom system consisting of
$\{\mi{Prop},\mi{MP},\Ver\}$.  The fact that  
the set of formulas known must be a subset of a downward closed set
is characterized by the following axiom:
\begin{namedaxiom}{DC}
\text{$\neg (K \phi_1 \land \ldots \land K \phi_m)$ if $\AX_{\Ver} \vdash 
\neg (\phi_1 \land \ldots \land\phi_n)$}.
\end{namedaxiom}
The key point here is that, as we shall show, a propositionally
consistent set of formulas that is
downward closed must be consistent with $\AX_{\Ver}$.  

The fact that the set of formulas that is known is k-compatible with a
downward closed set of formulas is characterized by the following axiom:
\begin{namedaxiom}{KC}
\begin{aligned}
& (K\phi_1 \land \ldots \land K\phi_n) \rimp (K\psi_1 \lor \ldots \lor
K\psi_m) \\
& \quad \text{if $\AX_{\Ver} 
\vdash \phi_1 \land \ldots \land  \phi_n \rimp (K\psi_1 \lor \ldots \lor
K\psi_m)$}.
\end{aligned}
\end{namedaxiom}
Axiom $\DC$ is just the special case of axiom $\KC$ where $m=0$.
It is also easy to see that $\KC$ (and therefore $\DC$) follow from
$\Ver$. 

Let $\AX_{\DC} = \{\mi{Prop},\mi{MP},\mi{DC}\}$ and let
$\AX_{\KC} = \{\mi{Prop},\mi{MP},\mi{KC}\}$.

\begin{theorem}\label{thm:completeness}
\mbox{}
\begin{itemize}
\item[(a)] $\AX_{DC}$ is a sound and complete 
axiomatization of $\LK$ with respect to KD45 
awareness structures; 
\item[(b)] $\AX_{KC}$ is a sound and complete 
axiomatization of $\LK$ with respect to 
KD45 impossible-worlds structures;
\item[(c)] $\AX_{\Ver}$ is a sound and complete 
axiomatization of $\LK$ with respect to 
S5 awareness structures and S5 impossible-worlds structures.
\end{itemize}
\end{theorem} 

\begin{proof}
We first prove soundness.  Consider axiom $\mi{DC}$.  Suppose that 
$\AX_{\Ver} \vdash \neg (\phi_1 \land \ldots \land\phi_n)$.  Let
$M = (W,W',\pi,\cA)$ be a KD45 awareness structure.  For each world $w'
\in W'$, it easily follows from Proposition~\ref{lem:closure} (taking $w
= w'$) that each instance of axiom $\Ver$ holds at $(M,w')$, as does
each instance of $\mi{Prop}$.  An easy argument by induction on the length
of proof then shows that, if $\AX_{\Ver} \vdash \psi$, then $(M,w')
\sat \psi$.  In particular, $(M,w') \sat \neg(\phi_1 \ldots \land
\phi_n)$.  It follows that, for each $w \in W$, we must have 
$(M,w) \sat \neg(K \phi_1 \land \ldots \land K \phi_n)$.
Essentially the same argument shows that axiom $\mi{DC}$ is sound in 
KD45 impossible-worlds structures.

A similar argument also shows the soundness of $\mi{KC}$ with respect to
KD45 impossible-worlds structures.  For suppose that
$M = (W,W',\pi,\cC)$ is an impossible-worlds structure, $w \in W$, 
$\AX_{\Ver} \vdash (\phi_1 \land \ldots \land\phi_n) \rimp (K\psi_1
\lor \ldots \lor K\psi_m)$, and
$(M,w) \sat K\phi_1 \land \ldots \land K\phi_n$.  Thus, 
$(M,w'') \sat \phi_1 \land \ldots \land \phi_n$ for all $w'' \in W'$.
But since  
each world in $W \cap W'$ is a model of $\AX_{\Ver}$, if $w' \in W
\cap W'$, we must have $(M,w') \sat K \psi_1 \lor \ldots \lor K
\psi_m$.  Moreover, since $W 
\cap W' \ne \emptyset$, 
there must be some world $w' \in W \cap W'$.  
It follows that, for some $j \in \{1, \ldots,
m\}$,  $(M,w') \sat K \psi_j$. 
Thus, $(M,w'') \sat \psi_j$ for all
$w'' \in W'$, so $(M,w) \sat K \psi_j$. It follows that $(M,w) \sat
K \psi_1 \lor \ldots \lor K \psi_m$, as desired.

Finally, as we have already observed, the soundness of $\Ver$ in S5
awareness and impossible-worlds structures follows easily from
Proposition~\ref{lem:closure}.  

For completeness, we start with part (a).  It suffices to show that,
given an $\AX_{\mi{DC}}$-consistent formula $\phi$, there exists a KD45
awareness structure $M$ and world $w$ such that $(M,w) \sat \phi$.  So
suppose that $\phi$ is $\AX_{\mi{DC}}$-consistent.  Let $G$ be a maximal
$\AX_{\mi{DC}}$-consistent set containing $\phi$.  
Let $F = \{\psi \mid K \psi \in G\}$.  We claim that
$F$ is $\AX_{\Ver}$-consistent.  If not, then there exists $\phi_1,
\ldots, \phi_n \in G$ such that $\AX_{\Ver} \vdash \neg (\phi_1
\land \ldots \land \phi_n)$.  But then by axiom $\mi{DC}$, 
we have that $\AX_{\mi{DC}} \vdash \neg(K \phi_1 \land \ldots \land K
\phi_n)$, contradicting the fact that $G$ is $\AX_{\mi{DC}}$-consistent.
Thus, $F$ is consistent with $\AX_{\Ver}$.  Let $F'$ be a maximal
$\AX_{\Ver}$-consistent set extending $F$.  Then it is easy to check
that $F'$ is a propositionally-consistent downward-closed set of
formulas that contains $F$.  Thus, by Theorem~\ref{t:awareness}, there
is a KD45 
awareness structure $M = (\{w,w'\}, \{w'\},\pi, \cA)$ such that $(M,w) \sat 
K \psi$ for all $\psi \in F$.  We can assume without loss of generality
that $w \ne w'$ and that $\pi(w)$ makes all the primitive propositions in
$F$ true.  (Note that this would not be the case if we were dealing with
S5 awareness structures.)  An easy induction on the structure of
formulas then shows that $(M,w) \sat \psi$ for all $\psi \in G$.
In particular, $(M,w) \sat \phi$.

For part (b), we 
use
much the same argument.  Suppose that $\phi$ is 
$\AX_{\mi{KC}}$-consistent.  Let $G$ be a maximal
$\AX_{\mi{KC}}$-consistent set containing $\phi$.   
Let $F = \{\psi \mid K \psi \in G\}$, and let 
$G' = F \cup  \{\neg \psi \mid \neg K \psi
\in G\}$.  We again claim that $G'$ is $\AX_{\Ver}$-consistent.  If
not, then there exists $K\phi_1, \ldots, K\phi_n, K \psi_1, \ldots, K
\psi_m \in G$ such that 
$\AX_{\Ver} \vdash (\phi_1 \land \ldots \land \phi_n) \rimp
(K\psi_1 \lor \ldots \lor K\psi_m)$.  By axiom $\mi{KC}$, we have that 
$\AX_{KC} \vdash (K\phi_1 \land \ldots \land K\phi_n) \rimp (K\psi_1
\lor \ldots \lor K\psi_m)$, 
contradicting the fact that $G$ is $\AX_{\mi{KC}}$-consistent.
Thus, $G'$ is consistent with $\AX_{\Ver}$.  Again, let $F'$ be a maximal
$\AX_{\Ver}$-consistent set extending $G'$.  Then it is easy to check
that $F'$ is a propositionally-consistent downward-closed set of
formulas that contains $F$; moreover the construction guarantees that
$F$ is k-compatible with $F'$.  Thus, by Theorem~\ref{t:awareness}, there
is a KD45 impossible-worlds
 structure $M = (\{w,w'\}, \{w',w''\}, \pi, \cC)$ such that
$(M,w) \sat  K \psi$ for all $\psi \in F$.  We can assume without loss
of generality 
that $w \ne w'$ and that $\pi(w)$ makes all the primitive propositions in
$F$ true.  An easy induction on the structure of
formulas then shows that $(M,w) \sat \psi$ for all $\psi \in G$.
In particular, $(M,w) \sat \phi$.

Finally, for part (c), let $\mbox{AX} = \{\mi{Prop},\mi{MP},\Ver\}$.
Suppose that $\phi$ is consistent with AX. Extend $\phi$ to a maximally
AX-consistent set $F$ of formulas.  It suffices to show that $F$ is
satisfiable in an S5 awareness structure and in an S5 impossible-worlds
structure.   In the case of awareness structures, consider the structure
$M = (\{w\}, \{w\}, \pi,\cA)$, where $\pi(w)(p) = \true$ iff $p \in F$
and $\cA(w) = \{\psi \mid K \psi \in F\}$.  We now show by induction on
the structure of formulas that $(M,w) \sat \psi$ iff $\psi \in F$.
If $\psi$ is a primitive proposition, then this is immediate from the
definition of $\pi$.  If $\psi$ has the form $\neg \psi'$, then the
result is immediate from the induction hypothesis.
If $\psi$ has the form $\psi_1 \land \psi_2$, this
is immediate from the observation that, since $F$ is a maximal
AX-consistent set and propositional reasoning is sound in AX that
$\psi_1 \land \psi_2 \in F$ iff $\psi_1 \in F$ and $\psi_2 \in F$.
If $\psi$ has the form $K\psi'$, note that if $K \psi' \in F$ then
$\psi' \in F$ (since $\Ver \in F$).  By the induction hypothesis,
$(M,w) \sat \psi'$.  Thus, $(M,w) \sat K\psi'$.  For the converse, if
$(M,w) \sat K \psi'$, suppose, by way of contradiction, that $K \psi'
\notin F$.  Then, by construction, $\psi' \notin \cA(w)\}$.  Thus,
$(M,w) \sat \neg K \psi'$, a contradiction.  

To show that $F$ is satisfiable in an S5 impossible-worlds structure,
consider the structure $M = (\{w\}, \{w,w'\}, \pi, \cC\}$, where
$\pi(w)(p) = \true$ iff $p \in F$ and $\cC(w') = \{\psi \mid K\psi \in
F\}$.  Thus, $\cC(w')$ is the same set of formulas as $\cA(w)$ in the
argument for S5 awareness structures.  An almost identical argument as
in the case of S5 awareness structures now
shows that $(M,w) \sat \psi$ iff $\psi \in F$.  We leave details to the
reader.  
\end{proof}

\begin{corollary}\label{cor:npcompleteness}
The satisfiability problem for 
the language $\LK$ with respect to KD45 awareness structures (resp., KD45
impossible-worlds structures, S5 awareness structures) is NP-complete.
\end{corollary}
\begin{proof}
NP-hardness follows immediately from the observation that 
$\LK$ contains propositional logic.
The fact that the
satisfiability 
problem with respect to each of these classes of structures is in NP follows from
the construction 
of Theorem~\ref{thm:completeness}, which shows that if a formula $\phi$ is
satisfiable
with respect to KD45 awareness structures (resp., KD45
impossible-worlds structures, S5 awareness structures), 
then it is consistent with respect to $\AX_{\mi{DC}}$ (resp.
$\AX_{\mi{KC}}$, $\AX_{\Ver}$), which in turn implies that it 
is satisfiable in a KD45 awareness structure (resp., KD45
impossible-worlds structure, S5 awareness structure) $M = (W, W',
\ldots)$ with two (resp., three, one) world(s).  Without loss of
generality, we can also assume 
that, in the case of awareness structures, at each world $w \in W$, 
$\cA(w)$ is a subset of $\Sub(\phi)$, the set of subformulas of
$\phi$, and $\pi(w)(p) = \true$ only if $p$ is a subformula of $\phi$;
similarly, in the case of impossible-worlds structures, we can assume
that for each impossible world $w'$, $\cC(w')$ is a subset of the
subformulas of $\phi$.  (If this is not true in $M$, then we can easily
modify $M$ so that this is true without affecting the truth of $\phi$ or
any subformula of $\phi$ in any world.)  Thus, we can guess a
satisfying structure for $\phi$ and verify that it satisfies $\phi$ in
time linear in the length of $\phi$.
\end{proof}

\section{Adding Probability}\label{sec:prob}

While the differences between K45, KD45$^-$, and KD45 impossible-worlds 
structures may appear minor, they turn out to be important when we add 
probability to the picture. 
As pointed out by Cozic \citeyear{r:cozic05}, standard models for
reasoning about probability suffer from the same logical omniscience
problem as models for knowledge. In the language considered by Fagin,
Halpern, and Megiddo \citeyear{FHM} (FHM from now on), there are
formulas that talk explicitly about probability.  A formula such as
$\ell(\Prime_n) = 1/3$ says that the probability that $n$ is prime is
$1/3$. 
In the FHM semantics, a probability is put on the set of worlds that
the agent considers possible.  The probability of a formula $\phi$ is then
the probability of the set of worlds where $\phi$ is true.  Clearly,
if $\phi$ and $\psi$ are logically equivalent, then
$\ell(\phi)=\ell(\psi)$ will be true.  However, the agent may not
recognize that $\phi$ and $\psi$ are equivalent, and so may not
recognize that $\ell(\phi)=\ell(\psi)$.  Problems of logical
omniscience with probability can to some extent be reduced to problems
of logical omniscience with knowledge in a logic that combines
knowledge and probability \cite{r:fagin94}.  For example, the fact
that an agent may not recognize $\ell(\phi)=\ell(\psi)$ when $\phi$
and $\psi$ are equivalent just amounts to saying that if $\phi \dimp
\psi$ is valid, then we do not necessarily want $K(\ell(\phi) =
\ell(\psi))$ to hold.  However, adding knowledge and awareness does
not prevent $\ell(\phi) = \ell(\psi)$ from holding.  This is not
really a problem if we interpret $\ell(\phi)$ as the objective
probability of $\phi$; if $\phi$ and $\psi$ are equivalent, it is an
objective fact about the world that their probabilities are equal, so
$\ell(\phi) = \ell(\psi)$ should hold.  On the other hand, if
$\ell(\phi)$ represents the agent's subjective view of the probability
of $\phi$, then we do not want to require $\ell(\phi) = \ell(\psi)$ to
hold.  This cannot be captured in all approaches.

To make this precise, we first clarify the logic we have in mind.
Let $\LKQU$ be  $\LK$ extended with linear inequality
formulas involving probability (called likelihood formulas), in the
style of FHM. 
A likelihood formula is of the form
$a_1\ell(\phi_1)+\dots+a_n\ell(\phi_n)\ge c$, where $a_1,\dots,a_n$
and $c$ are integers.  
(For ease of exposition, we restrict $\phi_1,\dots,\phi_n$ to be
propositional formulas in likelihood formulas; 
however, the techniques presented here can be extended to deal with
formulas that allow arbitrary nesting of $\ell$ and $K$).
We give semantics to these formulas by extending Kripke structures
with a probability distribution over the worlds 
that
the agent considers
possible. 
A \emph{probabilistic KD45 (resp., S5) Kripke structure} is a tuple
$(W,W',\pi,\mu)$, where $(W,W',\pi)$ is KD45 (resp., S5) Kripke
structure, and $\mu$ is a probability distribution over $W'$. 
To interpret likelihood formulas, we first define
$\event{\phi}{M}=\{w\in W\mid \pi(w)(\phi)=\true\}$, for
a propositional formula $\phi$. 
We then extend the semantics of $\LK$ with the following rule for
interpreting likelihood formulas:
\begin{itemize} 
\item[] $(M,w)\sat a_1\ell(\phi_1)+\dots+a_n\ell(\phi_n)\ge c$ iff 
$a_1 \mu(\event{\phi_1}{M}\cap W')+\dots+ a_n\mu(\event{\phi_n}{M}\cap
W')\ge c$.
\end{itemize} 
Note that the truth of a likelihood formula at a world does not depend
on that world; if a likelihood formula is true at a world of a
structure $M$, then it is true at every world of $M$.  

FHM give an axiomatization for likelihood formulas in probabilistic
structures. 
Aside from propositional reasoning axioms, one axiom captures
reasoning with linear inequalities.
A \emph{basic inequality formula} is a formula of the form 
$a_1 x_1 + \cdots + a_k x_k + a_{k+1} \le b_1 y_1 +\cdots + b_m y_m + b_{m+1}$, where $x_1,
\ldots, x_k, y_1, \ldots, y_m$ are (not necessarily distinct) variables.
A \emph{linear 
inequality formula} is a Boolean combination of basic linear inequality formulas.
A linear inequality formula
is valid if the resulting
inequality holds under every possible assignment of real numbers to
variables.  
For example, the formula $(2x + 3y \le 5 z) \land (x- y \le 12z) \rimp (3x +
2y \le 17z)$ is a valid linear inequality formula.
To get an instance of $\mi{Ineq}$, we replace each variable $x_i$
that occurs in a valid formula about linear inequalities by a
likelihood term of the form $\ell(\psi)$ (naturally, each
occurrence of the variable $x_i$ must be replaced by the same
primitive expectation term $\ell(\psi)$). 
(We can replace $\mi{Ineq}$ by a sound and complete axiomatization for
Boolean combinations of linear inequalities; one such axiomatization is
given in FHM.)

The other axioms of FHM are specific to probabilistic reasoning, and
capture the defining properties of probability distributions:
\begin{gather*}
 \ell(\truep) = 1\\
 \ell(\neg\phi) = 1-\ell(\phi)\\
 \ell(\phi\land \psi) + \ell(\phi\land\neg\psi) = \ell(\phi)
\end{gather*}

It is straightforward to extend all the approaches in
Section~\ref{sec:review} to the probabilistic setting.
In this section, we only consider  probabilistic awareness
structures and probabilistic impossible-worlds structures, because
the interpretation of both algorithmic knowledge and knowledge in
syntactic structures does not depend on the set of worlds or any
probability distribution over the set of worlds.

A \emph{KD45 (resp., S5) probabilistic awareness structure} is a tuple
$(W,W',\pi,\cA,\mu)$ where $(W,W',\pi,\cA)$ is a KD45 (resp., S5)
awareness structure and $\mu$ is a probability distribution over the
worlds in $W'$. Similarly, a \emph{KD45$^-$ (resp., KD45, S5)
probabilistic impossible-worlds structure} is a tuple
$(W,W',\pi,\cC,\mu)$ where $(W,W',\pi,\cC)$ is a KD45$^-$ (resp.,
KD45, S5) impossible-worlds structure and $\mu$ is a probability
distribution over the worlds in $W'$. 
Since the set of worlds that are assigned probability must be
nonempty, when dealing with probability, we must restrict to KD45
awareness structures and KD45$^-$ impossible-worlds structures,
extended with a probability distribution over the set of worlds the
agent considers possible. 
As we now show, adding probability to the language allows 
finer
distinctions between awareness structures and impossible-worlds
structures.

In probabilistic awareness structures, the axioms of probability
described by FHM are all valid.  For 
example, $\ell(\phi) = \ell(\psi)$  
is valid in probabilistic awareness structures if $\phi$ and $\psi$
are equivalent  
formulas.  
Using arguments similar to those in Theorem~\ref{t:awareness}, we can
show that $\neg K \neg \ell(\phi) = \ell(\psi)$  is valid in probabilistic
awareness structures.
Similarly, since $\ell(\phi) + \ell(\neg \phi ) = 1$ is valid
in probability structures,
 $\neg K(\neg(\ell(\phi) + \ell(\neg \phi ) = 1))$ is valid in 
probabilistic awareness structures. 

We can characterize properties of knowledge and likelihood in
probabilistic awareness structures axiomatically.
Let $\Prob$ denote a substitution instance of a valid formula in
probabilistic logic (using the FHM axiomatization).  By the observation
above, $\Prob$ is sound in probabilistic awareness structures.  Our 
reasoning has
to take this into account.  There is also an axiom $\KL$ that connects
knowledge and likelihood: 
\begin{namedaxiom}{KL}
K \phi \rimp \ell(\phi)>0.
\end{namedaxiom}

Let $\AX_{\Ver}^P$ denote the axiom system 
consisting of $\{\mi{Prop},\mi{MP},\Prob, \mi{KL},\Ver\}$.
Let $\DC^P$ be the following strengthening of $\DC$,
somewhat in the spirit of $\KC$:
\begin{namedaxiom}{$\DC^P$}
\begin{aligned}
& (K\phi_1 \land \ldots \land K\phi_n) \rimp (\psi_1 \lor \ldots \lor
\psi_m) \\
& \quad \text{if $\AX_{\Ver}^P 
\vdash \phi_1 \land \ldots \land  \phi_n \rimp (\psi_1 \lor \ldots \lor
\psi_m)$}\\
& \quad \text{and $\psi_1, \ldots, \psi_m$ are likelihood formulas.}
\end{aligned}
\end{namedaxiom}
Finally, even though $\Ver$ is not sound in KD45 probabilistic awareness
structures, a weaker version, restricted to likelihood formulas, is
sound, since there is a single probability distribution in probabilistic
awareness structures. Let $\WVer$ be the following axiom:
\begin{namedaxiom}{WVer}
K \phi \rimp \phi \text{ if $\phi$ is a likelihood formula}.
\end{namedaxiom}

Let
$\AX_{\DC}^P=\{\mi{Prop},\mi{MP},\mi{Prob},\mi{DC}^P,\WVer,\mi{KL}\}$
be the axiom system
obtained by replacing $\DC$ in
$\AX_{\DC}$ by $\DC^P$ and adding 
$\mi{Prob}$, 
$\WVer$, and $\mi{KL}$.

\begin{theorem}\label{thm:completenesspaw}\mbox{} 
\begin{enumerate} 
\item[(a)] 
$\AX_{\DC}^P$ is a sound and complete 
  axiomatization of $\LKQU$ with respect to 
KD45 probabilistic awareness  structures.
\item[(b)] 
$\AX_{\Ver}^P$ is a 
  sound and complete axiomatization of $\LKQU$ with respect to
S5 probabilistic awareness  structures.
\end{enumerate} 
\end{theorem} 
\begin{proof} 
We first prove soundness. 
We have already argued that $\mi{Prob}$ is sound in KD45
probabilistic awareness structures. 
It is easy to see that $\mi{KL}$ is sound: let $M=(W,W',\pi,\cA,\mu)$ be a KD45
probabilistic awareness structure, and let $w$ be a world in $W$ such
that $(M,w)\sat K\phi$. 
This means that $\phi$ is true at every world $w'\in W'$, and
therefore, $\mu(\event{\phi}{M}\cap W')=\mu(W')>0$, that is,
$(M,w)\sat\ell(\phi)>0$. 
Similarly, $\mi{WVer}$ is sound: let $M=(W,W',\pi,\cA,\mu)$ be a KD45
probabilistic awareness structure, and let $w$ be a world in $W$ such
that $(M,w)\sat K\phi$, with $\phi$ a likelihood formula. 
This means that $\phi$ is true at every world $w'\in W'$, and because
$\phi$ is a likelihood formula, the truth of $\phi$ does not depend on
the world.
Thus, if $\phi$ is true at some world, it is true at every
world; in particular, it is true at $w$, so that $(M,w)\sat\phi$, as
required. 
Finally, we show soundness of $\mi{DC}^P$, using an argument similar to
that in the proof of Theorem~\ref{thm:completeness}.
Suppose that $M = (W,W',\pi,\cA,\mu)$ is a KD45 probabilistic awareness
structure, $w \in W$, 
$\AX_{\Ver}^P \vdash (\phi_1 \land \ldots \land\phi_n) \rimp (\psi_1
\lor \ldots \lor \psi_m)$, for 
likelihood formulas $\psi_1,\dots,\psi_m$, and 
$(M,w) \sat K\phi_1 \land \ldots \land K\phi_n$.  Thus, 
$(M,w'') \sat \phi_1 \land \ldots \land \phi_n$ for all $w'' \in W'$.  
But since each world in $W'$ is a model of $\AX_{\Ver}^P$, if $w' \in
W'$, we must have $(M,w') \sat \psi_1 \lor \ldots \lor \psi_m$. 
Since $W' \ne \emptyset$, let $w'$ be an element of $W'$.  For some
$j \in \{1, \ldots, m\}$,  we must have $(M,w') \sat \psi_j$.
Because $\psi_j$ is a likelihood formula, and therefore its truth does
not depend on the world, if $\psi_j$ is true at some world, then
$\psi_j$ is true at every world.
In particular, $(M,w)\sat\psi_j$, and it follows that $(M,w) \sat 
\psi_1 \lor \ldots \lor \psi_m$, as desired.

The soundness of $\Ver$ in S5 probabilistic awareness structures
follows easily by induction on the structure of $\phi$ in $K\phi$,
using the fact that $\WVer$---the special case of $\Ver$ when $\phi$ is
a likelihood formula---is sound in probabilistic awareness structures,
and the argument for the soundness of $\Ver$ in S5 awareness
structures. 

For completeness, first consider part (a). Completeness
follows from combining techniques from the FHM completeness
proof with those of Theorem~\ref{thm:completeness}.  We briefly sketch the
main ideas here.  Define $\Sub_P(\phi)$ to be the least set containing
$\phi$, closed under subformulas, and containing $\ell(\psi) > 0$ if
it contains a propositional formula $\psi$.  It is easy to see that
$|\Sub_P(\phi)| \le 2|\phi|$. 
Suppose that $\phi$ is consistent with $\AX_{\DC}^P$.
Let $F$ be a maximal $\AX_{\DC}^P$-consistent subset of $\Sub_P(\phi)$
that includes $\phi$.  
Let $S$ consist of all truth assignments to primitive propositions.
Using techniques of FHM, we can show that there must be a probability 
measure $\mu$ on $S$ that makes all the likelihood formulas in $F$
true.  We remark for future reference that the FHM proof shows that we
can take the set of 
truth assignments which get positive probability to be polynomial in the
size of $|\phi|$, and we can assume that the probability is rational,
with a denominator whose size is polynomial in $|\phi|$.

Let $H = \{\psi \mid K \psi \in F\} \cup \{\psi\mid \psi \in F,
\text{$\psi$ is a likelihood formula}\}$.
Arguments almost identical to those in
Theorem~\ref{thm:completeness} show that $H$ must be
$\AX_{\DC}^P$-consistent.  Hence there is a maximal
$\AX_{\DC}^P$-consistent subset $F'$ of $\Sub_P(\phi)$ that contains $H$.
We now construct a KD45 awareness structure $(\{w\}\cup W',W',\cA,\mu')$ as follows.
There is a world $w_v$ in $W'$ corresponding to each truth assignment
$v$ such that $\mu(v) > 0$ and a world $w'$
corresponding to $F'$; we define $\mu'$ on $W'$ so that $\mu'(w') = 0$
and $\mu'(w_v) = \mu(v)$.  Define $\pi$ so that $\pi(w_v) = v$,
$\pi(w)(p) = \true$ iff $p \in F$ and $\pi(w')(p) = \true$ iff $p \in
F'$.  Finally, define $\cA$ so that $\cA(w_v) = \emptyset$, $\cA(w') =
\{\psi\mid K \psi \in F'\}$ and $\cA(w) = \{\psi\mid K \psi \in F\}$.  Now the
same ideas as in the proof of Theorem~\ref{thm:completeness} show that,
for each formula $\psi \in \Sub_P(\phi)$ we have that 
$(M,w') \sat \psi$ iff $\psi \in F'$ and $(M,w) \sat \psi$ iff $\psi \in
F$.  Thus, $(M,w) \sat \phi$.

The proof of completeness for part (b) is similar in spirit; the
modifications required are exactly those needed to prove
Theorem~\ref{thm:completeness}(c).  We leave details to the reader.
\end{proof}

Things change significantly when we move to probabilistic impossible-worlds
structures.  In particular, $\Prob$ is no longer sound.  For example,
even if $\phi \dimp \psi$ is valid, $\ell(\phi) = \ell(\psi)$ is not
valid, because we can have an impossible possible world with positive
probability where both $\phi$ and $\neg \psi$ are true.  Similarly, $\ell(\phi)
+ \ell(\neg \phi) = 1$ is not valid.  Indeed, both $\ell(\phi) +
\ell(\neg \phi) > 1$ and $\ell(\phi) + \ell(\neg \phi) < 1$ are both
satisfiable in impossible-worlds structures: the former requires that
there be an impossible possible world that gets positive probability
where both $\phi$ and $\neg \phi$ are true, while the latter requires an
impossible possible world with positive probability where neither is true. 
As a consequence, 
it is not hard to show that both 
$K \neg (\ell(\phi) = 
\ell(\psi))$  
and $K(\neg (\ell(\phi) + \ell(\neg \phi) = 1))$ are 
satisfiable in such impossible-worlds structures.%
\footnote{We remark that 
Cozic~\citeyear{r:cozic05}, who  
considers the logical omniscience problem in the context of 
probabilistic reasoning, 
makes somewhat similar points.  
Although he does not formalize things quite the way we do, he observes 
that, in his setting, impossible-worlds structures seem more 
expressive than awareness structures.} 
In fact,  
the only constraint on  
probability in probabilistic impossible-worlds structures is 
that it must be between 0 and 1.  This constraint is expressed by the 
following axiom $\mi{Bound}$: 
\begin{namedaxiom}{Bound}
\ell(\phi)\ge 0\land \ell(\phi)\le 1.
\end{namedaxiom}

We can characterize properties of knowledge and
likelihood in probabilistic impossible-worlds structures axiomatically.
Let $\AX^B_\imp = \{\mi{Prop},\mi{MP}, \mi{Ineq}, \mi{Bound}, \KL, \WVer\}$.
We can think of $\AX^B_\imp$ as being the core of probabilistic reasoning in
impossible-worlds structures. 

Let $\AX_{\Ver}^{B}$ denote the axiom system 
consisting of $\{\mi{Prop},\mi{MP}, \mi{Ineq}, \mi{Bound},\Ver,
\mi{KL}\}$.  
Let $\KC^P$ denote the following extension of $\KC$:
\begin{namedaxiom}{$\KC^P$}
\begin{aligned}
& (K\phi_1 \land \ldots \land K\phi_n) \rimp (\psi_1 \lor \ldots \lor
\psi_m)\\
& \quad\text{if $\AX_{\Ver}^{P} 
\vdash \phi_1 \land \ldots \land  \phi_n \rimp (\psi_1 \lor \ldots \lor
\psi_m)$}\\
& \quad\text{and $\psi_j$ is either 
a
likelihood formula or of the form $K\psi'$,
for $j = 1, \ldots, m$.}
\end{aligned}
\end{namedaxiom}
Here again, $\mi{DC}^P$ is a special case of $\mi{KC}^P$. 
Let $\AX_{\KC}^B=\{\mi{Prop},\mi{MP},\mi{Bound},\mi{KC}^P,\WVer,\mi{KL}\}$ 
obtained by replacing $\KC$ in $\AX_{\KC}$ by $\KC^P$ 
and adding $\mi{Bound}$, $\WVer$ and $\mi{KL}$.

\begin{theorem}\label{thm:completenesspimp}   \mbox{}
\begin{enumerate} 
\item[(a)] 
$\AX^B_\imp$ is a sound
and complete  
  axiomatization of $\LKQU$ with respect to KD45$^{-}$ probabilistic
impossible-worlds structures.
\item[(b)] 
$\AX_{KC}^B$ is a sound and complete 
  axiomatization of $\LKQU$ with respect to KD45 probabilistic
impossible-worlds structures.
\item[(c)] 
$\AX_{\Ver}^B$ is a 
  sound and complete axiomatization of $\LKQU$ with respect to 
  S5 probabilistic impossible-worlds structures with 
  probabilities. 
\end{enumerate} 
\end{theorem} 

\begin{proof}
We first prove soundness.
The argument is similar to the argument for soundness in
Theorem~\ref{thm:completenesspaw}. 
That $\mi{KL}$ and $\WVer$ are sound in probabilistic impossible-worlds
structures follows from the same argument as in
Theorem~\ref{thm:completenesspaw}. 
To show that $\mi{Bound}$ is sound, note that for any probabilistic
impossible-worlds structure $M$, $\event{\phi}{M}\cap W'\subseteq W'$,
so that $0\le \mu(\event{\phi}{M})\le 1$. 
Because this is independent of the actual world, $(M,w)\sat
\ell(\phi)\ge 0 \land \ell(\phi)\le 1$ holds.

We show soundness of $\mi{KC}^P$ with respect to
KD45 probabilistic impossible-worlds structures.  For suppose that
$M = (W,W',\pi,\cC,\pi)$ is a KD45 probabilistic impossible-worlds
structure, $w \in W$,  
$\AX_{\Ver} \vdash (\phi_1 \land \ldots \land\phi_n) \rimp (\psi_1
\lor \ldots \lor \psi_m)$, where each $\psi_j$ either a likelihood
formula or of the form $K\psi'$, and
$(M,w) \sat K\phi_1 \land \ldots \land K\phi_n$.  
Thus, $(M,w'') \sat \phi_1 \land \ldots \land \phi_n$ for all $w'' \in
W'$.  
But since each world in $W \cap W'$ is a model of $\AX_{\Ver}^P$, if
$w' \in W \cap W'$, we must have $(M,w') \sat \psi_1 \lor \ldots
\lor \psi_m$.  
Moreover, since $W \cap W' \ne \emptyset$, there must be some world
$w'\in W\cap W'$. 
It follows that, for some
$j \in \{1, \ldots, m\}$, $(M,w') \sat \psi_j$. 
There are two cases.
If $\psi_j$ is a likelihood formula, then its truth does
not depend on the world, so that if $\psi_j$ is true at some world,
then $\psi_j$ is true at every world.
In particular, $(M,w)\sat\psi_j$, and it follows that $(M,w) \sat 
\psi_1 \lor \ldots \lor \psi_m$, as desired.
If $\psi_j$ is a formula of the form $K\psi'$, then
$(M,w'') \sat \psi'$ for all $w'' \in W'$, so $(M,w) \sat K
\psi'$, that is, $(M,w)\sat \psi_j$. 
It follows that $(M,w) \sat \psi_1 \lor \ldots \lor \psi_m$, as
desired.

Finally, as in the proof of Theorem~\ref{thm:completenesspaw}, the
soundness of $\Ver$ in S5 probabilistic impossible-worlds structures 
follows by induction on the structure of $\phi$ in $K\phi$.

For completeness, we prove part (a).  Given a formula $\phi$ consistent
with $\AX_\imp$, 
let $F$ be a maximal $\AX_\imp$-consistent subset of $\Sub_P(\phi)$
that includes $\phi$.  Consider the basic likelihood formulas in $F$.
>From these, we can get a system of linear inequalities 
by replacing each term $\ell(\psi)$ by a variable $x_\psi$.  We add
an inequality $0 \le x_\psi \le 1$ for each formula $\psi \in
\Sub_P(\phi)$.  Using the arguments of FHM, we can show that this set of
inequalities must be satisfiable (otherwise $F$ would not be $\AX_\imp$
consistent.)  Take a solution.  Without loss of generality, 
we have subformulas listed so that $x_{\psi_1} \le x_{\psi_2} \le \ldots
\le x_{\psi_n}$.  Let $n^*$ be the least $m$ such that $x_{\psi_m} = 1$;
if $x_{\psi_n} < 1$, then let $n^* = n+1$.
Consider a probabilistic impossible-worlds structure
$(\{w\}, \{w_1, \ldots, w_{n+1},w\}, \pi,
\cC, \mu)$, where  
we define $\pi$, $\cC$ and $\mu$ as follows:
\begin{itemize}
\item $\pi(w)(p) = \true$ iff $p \in F$; 
\item $\mu(w_1) = x_{\psi_1}$, $\mu(w_j) = x_{\psi_{j}} -
x_{\psi_{j-1}}$ for $j = 2, \ldots, n$, and $\mu(w_{n+1}) = 1 - \mu(w_n)$;
\item $\cC(w_j) = \{\psi_j, \ldots, \psi_n\}$ for $j = 1, \ldots, n^*$
\item $\cC(w_{j}) = \cC(w_{n^*})$ if $j = n^* + 1, \ldots, n+1$.
\end{itemize}
We leave it to the reader to show that $(M,w) \sat \phi$.
The proof for parts (b) and (c) is similar in spirit and left to the
reader. 
\end{proof}

Observe that Theorem~\ref{thm:completenesspimp} is true even though  
probabilities are standard in impossible worlds:  
the probabilities of worlds still sum to 1.  It is just the truth 
assignment to formulas that behaves in a nonstandard way in impossible 
worlds.  
Intuitively, while the awareness approach is modeling certain
consequences of resource-boundedness in the context of knowledge, it
does not do so for probability.  On the other hand, the
impossible-worlds approach seems to extend more naturally to accommodate
the consequences of resource-boundedness in probabilistic reasoning;
see Section~\ref{sec:pragmatics} for more discussion of this issue.

\begin{corollary}
The satisfiability problem for 
the language $\LKQU$ with respect to 
KD45 probabilistic awareness structures (resp., S5 probabilistic
awareness structures, KD45$^-$ 
probabilistic impossible-worlds structures, KD45 probabilistic
impossible worlds structures, S5 probabilistic impossible-worlds
structures) 
is NP-complete. 
\end{corollary}
\begin{proof}
Again, NP-hardness follows immediately from the observation that 
$\LKQU$ contains propositional logic.  
The fact that the satisfiability problem 
with respect to each of these classes of structures is
in NP follows from the constructions of
Theorems~\ref{thm:completenesspaw} and \ref{thm:completenesspimp},
which show that if a formula $\phi$ is satisfiable 
with respect to 
KD45 probabilistic awareness structures (resp., S5 probabilistic
awareness structures, KD45$^-$ 
probabilistic impossible-worlds structures, KD45 probabilistic
impossible worlds structures, S5 probabilistic impossible-worlds
structures), 
then it is
consistent with respect to $\AX_{\mi{DC}}^P$ (resp., $\AX_{\Ver}^P$,
$\mi{AX}_\mi{imp}^B$, $\mi{AX}^B_{\KC}$, $\mi{AX}^B_{\Ver}$) which in
turn implies that it is satisfiable in a KD45 probabilistic awareness
structure (resp., S5 probabilistic awareness structure, KD45$^-$
probabilistic impossible-worlds structure, KD45 probabilistic
impossible worlds structure, S5 probabilistic impossible-worlds
structure) $M = (W, W', \ldots)$ with a small number of
worlds---polynomial in the length of $\phi$ in each case. 
Just like in the proof of Corollary~\ref{cor:npcompleteness}, without
loss of generality, we can assume that, in the case of
probabilistic awareness structures, at each world $w \in W$, $\cA(w)$
is a subset of $\Sub(\phi)$, the set of subformulas of $\phi$, and
$\pi(w)(p) = \true$ only if $p$ is a subformula of $\phi$; similarly,
in the case of probabilistic impossible-worlds structures, we can
assume that for each impossible world $w'$, $\cC(w')$ is a subset of
the subformulas of $\phi$.  
Finally, using the arguments of FHM, we can argue without loss of
generality that the probability distributions $\mu$ are described in
size polynomial in the length of $\phi$. (The probability
distributions in all structures can be taken to assign
small---polynomial-size---rational probabilities to every world, where
the size of a rational number is the sum of the sizes of the numerator
and denominator when they are relatively prime.)
Thus, we can guess a satisfying structure for $\phi$ and verify that
it satisfies $\phi$ in time polynomial in the length of $\phi$.
\end{proof}

\section{Pragmatic Issues}\label{sec:pragmatics} 

Even in settings where the four approaches are equi-expressive, they 
model lack of logical omniscience quite differently. 
We thus have to deal with different issues when attempting to use one of 
them in practice.  For example, 
if we are using a syntactic structure to represent a given 
situation, we need to explain where the function $\cC$ is coming from; 
with an awareness structure, we must 
explain where the awareness function is coming from; 
with an algorithmic knowledge structure, we must explain where the 
algorithm is coming from; and 
with an impossible-worlds structure, we must explain what the 
impossible worlds are.  

There seem to be three quite 
distinct intuitions underlying the lack of logical omniscience 
As we now discuss, these intuitions can guide the choice of approach, 
and match closely the solutions described above.  
We discuss, for each intuition, the extent to which each of the
approaches to dealing with logical omniscience can capture that
intuition.
While the discussion in this section is somewhat informal, we believe
that these observations will prove important when actually trying to
decide how to model lack of logical omniscience in practice.

\subbit{Lack of Awareness}\label{sec:lackofawareness}
The first intuition is 
lack of awareness of some primitive notions: for example, when 
trying to consider possible outcomes of an attack on Iraq, the worlds 
can be taken to represent the outcomes.  An agent simply may be unable 
to contemplate some of the outcomes of an attack, so cannot consider 
them possible, let alone know that they will happen or not happen.   
This can be modeled reasonably well using an awareness structure where the 
awareness function is \emph{generated by primitive propositions}.   
We assume that the agent is unaware of certain primitive propositions,  
and is unaware of 
exactly those formulas that contain a primitive proposition of which the 
agent is unaware.  This intuition is quite prevalent in the economics 
community, and all the standard approaches to modeling lack of logical 
omniscience in the economics literature \cite{MR94,MR99,DLR98,HMS03} 
can essentially be understood in terms of awareness structures where 
awareness is generated by primitive propositions \cite{Hal34,HR05}. 

If awareness is generated by primitive propositions, constructing an 
awareness structure corresponding to a particular situation is no more 
(or less!) complicated that constructing a Kripke structure to capture 
knowledge without awareness.   
Determining the awareness sets for notions of awareness that are not 
generated by primitive propositions may be more complicated.  
It is also worth stressing that an awareness 
structure must be understood as the modeler's view of the situation.
For example, if awareness is generated by primitive  
propositions and agent 1 is not aware of a primitive proposition $p$, 
then agent 1 cannot contemplate a world where $p$ is true (or false); in 
the model from agent 1's point of view, there is no proposition $p$.

How do the other approaches fare 
in modeling lack of awareness?
To construct a syntactic structure, we need to know all sentences that 
an agent knows before constructing the model. 
This may or may not be reasonable.  
But it does not help one discover properties of knowledge in a 
given situation. 
As observed in \cite{r:fagin95}, the syntactic approach is really only a
representation of knowledge.
Algorithmic knowledge can deal with lack of awareness reasonably well, 
provided that there is an algorithm $\A_a$ for determining what the
agent is aware of and an algorithm $\A_k$ for determining whether a
formula is true in every world in $W'$, the set of worlds that the agent
considers possible.  
If so, given a query $\phi$, the algorithmic approach would simply 
invoke
$\A_a$ to check whether the agent is aware of $\phi$; if so, then the
agent invokes $\A_k$.  
For example, if awareness is generated by primitive propositions, then 
$\A_a$ is the algorithm that, given query $\phi$, checks whether all the
primitive propositions in $\phi$ are ones the agent is aware of; and we
can take $\A_k$ to be the algorithm that does model checking 
to see if $\phi$ is true in every world of $W'$. (This can be done in
time polynomial in $W'$; see \cite{r:fagin95}.)
In impossible-worlds structures, we can interpret lack of awareness of
$\phi$ as meaning that neither $\phi$ nor $\neg \phi$ is true at all
worlds the agent considers possible.   Thus, if there is any nontrivial
lack of awareness, then all the worlds that the agent considers possible
will be impossible worlds.  However, these impossible worlds have a
great deal of structure: we can require that for all the formulas $\phi$
that the agent is aware of, exactly one of $\phi$ and $\neg \phi$ is
true at each world the agent considers possible.  As we observed
earlier, an awareness structure must be viewed as the \emph{modeler's}
view of the situation.  Arguably, the impossible-worlds structure
better captures the agent's view.

\subbit{Lack of Computational Ability} The second intuition is 
computational: an agent simply might not have the resources to compute 
the required answer.  But then the question is how 
to model this lack of computational ability.  
There are two cases of interest, depending on whether we have an 
 explicit algorithm in mind. 
If we have an explicit algorithm, then it is relatively 
straightforward. 
For example, Konolige \citeyear{Kon1} uses a syntactic approach and 
gives an explicit characterization of $\cC$ by taking it to be the set 
of formulas that can be derived from a fixed initial set of formulas by 
using a sound but possibly incomplete set of inference rules.   
Note that Konolige's approach makes syntactic knowledge an instance of 
algorithmic knowledge.  
(See also Pucella \citeyear{r:pucella06c} for more details on 
knowledge algorithms given by inference rules.) 

Algorithmic knowledge can be viewed as a generalization of Konolige's
approach in this setting, since it allows for the possibility that the
algorithm used by the agent to compute what he knows  may not be easily
expressible as a set of inference rules over formulas.  
For example, Berman, Garay, and Perry \citeyear{r:berman89} implicitly use a 
particular form of algorithmic knowledge in their analysis of  
\emph{Byzantine agreement} (this is the problem of getting all nonfaulty 
processes in a system to coordinate, despite the presence of failures). 
Roughly speaking, they allow agents to perform limited tests based on 
the information they have; agents know only what follows from these 
limited tests.  But these tests are not characterized axiomatically. 
As shown by Halpern and Pucella \citeyear{r:halpern02e}, algorithmic 
knowledge is also a natural way to capture adversaries in security protocols. 

\begin{example} Security protocols are generally analyzed in the 
presence of an adversary that has certain capabilities for decoding 
the messages he intercepts.    
There are of course restrictions on the capabilities of a reasonable 
adversary.  
For instance, the adversary may not explicitly know that he has a 
given message if that message is encrypted using a key that the 
adversary does not know.  
To capture these restrictions, Dolev and Yao \citeyear{r:dolev83} gave 
a now-standard description of the capabilities of adversaries.   
Roughly speaking, a Dolev-Yao adversary can decompose messages, or 
decipher them if he knows the right keys, but cannot otherwise 
``crack'' encrypted messages.  
The adversary can also construct new messages by concatenating known 
messages, or encrypting them with a known encryption key.   

Algorithmic knowledge is a natural way to capture the knowledge of a 
Dolev-Yao adversary \cite{r:halpern02e}.  
We can use a knowledge algorithm  
$\ADY$ to compute whether the adversary can \emph{extract} a message $\msg$ 
from a set $H$ of messages that he has intercepted, where the extraction 
relation $H\derivesDY \msg$ is defined by following inference rules: 
\[ \begin{array}{cccc} 
\Rule{\msg\in H}{H\derivesDY \msg} &  
 \Rule{H\derivesDY\encr{\msg}{\ky} \quad H\derivesDY 
   \ky}{H\derivesDY \msg} & %
\Rule{H\derivesDY \msg_1\cdot \msg_2}{H\derivesDY \msg_1} &  
 \Rule{H\derivesDY \msg_1\cdot \msg_2}{H\derivesDY \msg_2}, 
\end{array}\] 
where $\msg_1\cdot\msg_2$ is the concatenation of messages 
$\msg_1$ and $\msg_2$, and $\encr{\msg}{\ky}$ is the encryption of 
message $\msg$ with key $\ky$.  

The knowledge algorithm  
$\ADY$  
simply implements a search  
for the derivation of a message $\msg$ from the messages  
that the adversary has received and the initial set of keys, using the 
inference rules above. 
More precisely, we assume the language has formulas $\mi{has}(\msg)$, 
interpreted as ``the agent possesses message $\msg$''. When queried 
for a formula $\mi{has}(\msg)$, the knowledge algorithm $\ADY$ simply 
checks if $H\derivesDY \msg$, where $H$ is the set of messages 
intercepted by the adversary.  
Thus, the formula $K(\mi{has}(\msg))$, which is true if and only if 
$\ADY$ says $\text{``Yes''}$ to query $\mi{has}(\msg)$, that is, if 
and only if $H\derivesDY \msg$, says that the adversary can extract 
$\msg$ from the messages he has intercepted.  
\end{example}

However, 
even when our intuition is computational, at times the details of the 
algorithm do not matter (and, indeed, may not be known to the 
modeler). 
In this case, awareness may be more useful than algorithmic knowledge. 
\begin{example}\label{xam:Alice+Eve} 
Suppose that Alice is trying to reason about whether or not an 
eavesdropper Eve has managed to decrypt a certain message.  The intuition 
behind Eve's inability to decrypt is computational, but Alice does not 
know which algorithm Eve is using.  An algorithmic 
knowledge 
structure is typically appropriate if there are only a few algorithms that 
Eve might be using, and her ability to decrypt depends on the 
algorithm.%
\footnote{What is required here is an algorithmic knowledge  
structure with two agents.  There will then be different algorithms 
for Eve associated with different states.  We omit here the 
straightforward details of how this  
can be done; see \cite{r:halpern94}.} 
On the other hand, Alice might have no idea of what Eve's algorithm is, 
and might not care.  All that matters to her analysis is whether Eve has 
managed to decrypt.  In this case, using a syntactic structure or an 
awareness structure seems more appropriate.   
Suppose that Alice wants to model her uncertainty regarding whether Eva 
has decrypted the message.  She could  then use an awareness structure 
with some possible worlds where Eve is aware of the message, and others 
where she is not, with the appropriate probability on each set. 
Alice can then reason about the likelihood that Eve has decrypted the 
message without worrying about how she decrypted it. 
\end{example}

What about the impossible-worlds approach?  It cannot directly represent
an algorithm, of course.  However, if there is algorithm $\A$ that
characterizes an agent's computational process, then we can simply take
$W' = \{w'\}$ and define $\cC(w') = \{\phi \mid \A(\phi) =
\text{``Yes''}\}$.  Indeed, we can give a general computational
interpretation of the impossible-worlds approach.  The worlds $w$ such
that $\cC(w)$ are precisely those worlds where the algorithm answers
``Yes'' when asked about $\phi$.  If neither $\phi$ nor $\neg \phi$ is
in $\cC(w)$, that just means that the algorithm was not able to
determine whether $\phi$ was true or false.  If the algorithm answers
``Yes'' to both $\phi$ and $\neg \phi$, then clearly the algorithm is
not sound, but it may nevertheless describe how a resource-bounded agent
works.  

This intuition also suggests how we can model the lack of computational
ability illustrated by Example~\ref{xam:Alice+Eve} using impossible
worlds. 
If $\it{cont}(m) = \phi$ is
the statement that the content of the message $m$ is $\phi$, then in a world
where Alice cannot decrypt $\phi$, neither $\it{cont}(m) = \phi$ and
$\neg(\it{cont}(m) = \phi)$ would be true.  

\subbit{Imperfect Understanding of the Model} 
Sometimes an agent's lack of logical omniscience is best thought of as 
stemming from ``mistakes'' in constructing the model (which perhaps are 
due to lack of computational ability). 

\begin{example}\label{xam:primality} 
Suppose that Alice does not know whether a number $n$ is prime. 
Although her ignorance regarding $n$'s primality can be viewed as 
computationally based---given enough time and energy, she could 
in principle figure out whether $n$ is prime---she is not using a 
particular algorithm to compute her knowledge (at least, not one that 
can be easily described).  Nor can her state of mind be modeled in a 
natural way using an awareness structure or a syntactic structure. 
Intuitively, there should at least two worlds she considers possible, 
one where $n$ is prime, and one where $n$ is not.   
However, $n$ is either prime or it is not. 
If $n$ is actually prime, then there cannot be a possible world 
where $n$ is not prime; similarly, if $n$ is composite, there cannot be a 
possible world where $n$ is prime.  This problem can be modeled 
naturally using impossible worlds.  Now there is no problem 
having a world where $n$ is prime (which is an impossible 
world if $n$ is actually composite) and a world where $n$ is 
composite (which is an impossible world if $n$ is actually 
prime). 
In this structure, it is also seems reasonable to assume that 
Alice knows that she does not know that $n$ is prime (so that the 
formula $\neg K \Prime_n$ is true even in the impossible worlds). 

It is instructive to compare this with the awareness approach.  Suppose 
that $n$ is indeed prime and an external modeler knows this.  Then he 
can describe Alice's state of mind with one world, where $n$ is prime, 
but Alice is not aware that $n$ is prime.  Thus, $\neg K \Prime_n$ holds 
at this one world.  But note that this is not because Alice considers it 
possible that $n$ is not prime; rather, it is because Alice cannot 
compute whether $n$ is prime.  If Alice is aware of the formula $\neg K 
\Prime_n$ at this one world, then $K \neg K \Prime_n$ also holds. 
Again, we should interpret this as saying that Alice knows that she 
cannot compute whether $n$ is prime. 
\end{example} 

The impossible-worlds approach seems like a natural one in  
Example~\ref{xam:primality} and many other settings.   
As we saw, awareness in this situation does not quite capture 
what is going on here.
Algorithmic knowledge fares somewhat better, but it would require us to have 
a specific algorithm in mind; in Example~\ref{xam:primality}, this 
would force us to interpret ``knows that a number is prime'' as 
``knows that a number is prime as tested by a particular factorization 
algorithm''.  
The impossible-worlds approach can sometimes be difficult to apply,
however,  because it is  
not always clear what impossible worlds to take. 
While  
there has been a great deal of discussion (particularly 
in the philosophy literature) concerning the 
 ``metaphysical status'' of impossible worlds (cf.~\cite{r:stalnaker96}), 
the pragmatics of generating 
impossible worlds has received comparatively little 
attention. Hintikka \citeyear{r:hintikka75} argues that  
Rantala's \citeyear{r:rantala75} urn models 
are suitable candidates for impossible worlds.  
In decision theory, Lipman \citeyear{r:lipman99} uses 
impossible-worlds structures to represent the preferences of an agent 
who may not be able to distinguish logically equivalent outcomes; the 
impossible worlds are determined by the preference order. 
None of these approaches address the problem of 
generating the impossible worlds even in a simple example such as 
Example~\ref{xam:primality}, especially if the worlds have some 
structure.   
We view impossible worlds as describing the agent's subjective 
view of a situation. 
The modeler may know that these impossible worlds are truly 
impossible, but the agent does not.   
In many cases, the intuitive reason that the agent does not realize 
that the impossible worlds are in fact impossible is that the agent 
does not look carefully at the worlds.   
Consider Example~\ref{xam:primality}.  
Let $\Prime_n$, for various choices of $n$, be a primitive proposition 
saying that the number $n$ is prime.  
Suppose that the worlds are models of arithmetic, which include as domain 
elements the natural numbers with multiplication defined on them.   
If $\Prime_n$ is interpreted as being true in a world when there do 
not exist numbers $n_1$ and $n_2$ in that world such that $n_1 \times 
n_2 = n$, then how does the agent conceive of the impossible worlds?   
If the agent were to look carefully at 
a world where $\Prime_n$ holds, he might realize that there are in 
fact two numbers $n_1$ and $n_2$ such that $n_1 \times n_2 = n$.   
But if $n$ is not prime, how do we capture the fact that the agent 
``mistakenly'' constructed a world where there are numbers $n_1$ and 
$n_2$ such that $n_1 \times n_2 = n$ if we also assume that the agent 
understands basic multiplication?

We now sketch a new approach to constructing an impossible-worlds 
structure that seems appropriate for such examples. 
The approach is motivated by the observation that 
the set of worlds  
in a Kripke structure 
is explicitly  
specified, as is the truth assignment on these worlds.   
Introspectively, this is not the way in which we model situations.  
Instead, the set of possible worlds is described 
implicitly, as is the interpretation $\pi$, 
as the set of worlds satisfying some condition.%
\footnote{In multiagent settings, where the worlds that the agent 
considers possible are defined by an accessibility 
relation, we expect the accessibility relation  to be described 
implicitly as well.}  
This set of worlds may well include some impossible worlds. 
The impossible-worlds structure corresponding to a situation, 
therefore, is made up of all worlds satisfying the implicit 
description, perhaps refined so that ``clearly impossible'' worlds are 
not considered. What makes a world clearly impossible should be 
determined by a simple test; for example,  
such a simple test might determine that 3 is prime, but would not be 
able to determine that $2^{24036583} - 1$ is prime. 

We can formalize this construction as follows.  
An implicit structure is a tuple $I=(S,T,\cC)$, where $S$ is a set
of possible worlds, $T$ is a filter on worlds (a test on
worlds that returns either $\true$ or $\false$), and $\cC$ associates
with every world in $S$ a set (possibly inconsistent) of propositional
formulas. 
Test $T$ returns $\true$ for every world in $S$ that the agent
considers possible.
An implicit structure $I=(S,T,\cC)$ induces an impossible-worlds 
structure $M_I=(W,W',\pi,\cC)$ given by:
\begin{align*} 
W & = \{ w\in S \mid \text{$\cC(w)$ is consistent}\}\\
W' & = \{ w\in S\mid T(w)=\true\}\\
\pi(w) &= \cC(w)|_{\Phi} \quad \text{for $w \in W$}\\
\cC &= \cC|_{(W'-W)}.
\end{align*} 
We can refine the induced impossible-worlds structure by alotting more
resources to test $T$.
Intuitively, as an agent performs more introspection, she can
recognize more worlds as being impossible.
(Manne \citeyear{r:manne05} investigates a related approach, using a
temporal structure at each world to capture the evolution of knowledge
as the agent introspects over time.)  

Consider the primality example again.   
The agent is likely to care about the primality of only a few numbers, 
say $n_1, \ldots, n_k$.   
Let $\Phi = \{\Prime_{n_1}, \ldots, \Prime_{n_k}\}$.  
The agent's inability to compute whether $n_1, \ldots, n_k$ are 
prime is described implicitly by having worlds where any combination 
of them is prime. 
The details of how multiplication works in a world is not 
specified in the implicit description. 
Thus, the implicit structure $I=(S,T,\cC)$ corresponding to this
description will have $S$ consisting of $2^k$ worlds, where each world
is a standard model of arithmetic together with a truth assignment to
the primitive  propositions in $\Phi$. 
The set of formulas $\cC(w)$ consists of all propositional formulas
true under the truth assignment at $w$. 
The agent realizes that all but one of these worlds is impossible, but 
cannot compute which one is the possible world. 
Thus, we take $T(w)=\true$ for all worlds $w$. 
Of course, after doing some computation, the agent may realize that, 
say, $n_1$ is prime and $n_2$ is composite.   
The agent would then refine the model by taking $T$ to consider
possible only worlds in which $n_1$ is prime and $n_2$ is composite. 

The use of an implicit description as a recipe for constructing possible 
(and impossible) worlds is quite general, 
as the following example illustrates. 
\begin{example}\label{xam:policies} 
Suppose that we have a database of implications: rules of the form  
$C_1\rimp C_2$, where $C_1$ and $C_2$ are conjunctions of 
literals---primitive propositions and their negation.  
Suppose that the vocabulary of the conclusions of these rules is 
disjoint from the vocabulary of the antecedents.  
This is a slight simplification of, for example, digital rights 
management policies, where the conclusion typically has the form 
\emph{Permitted(a,b)} or $\neg$\emph{Permitted(a,b)} for some agent $a$ 
and action $b$, and \emph{Permitted} is not allowed to appear in the 
antecedent of rules \cite{HW03}. 
Rather than explicitly constructing the worlds compatible with the 
rules, a user might construct a naive implicit description of  
them. 
More specifically, suppose that we have a finite set of agents,  
say $a_1, \ldots, a_n$, and a finite set of actions, say  
$b_1,\ldots, b_m$. 
Consider the implicit structure $I=(S,T,\cC)$, where each world $w$ in
$S$ is a truth assignment to the atomic
formulas that appear in the antecedents of rules, augmented with all the 
literals in the conclusions of rules whose antecedent is true in $w$;
furthermore, take $T(w)=\true$ for all $w\in S$, and $\cC(w)$ to be
all propositional formulas true under the truth assignment at world
$w$. 
Thus, for example, if a rule says $\mi{Student}(a) \land \mi{Female}(a) 
\rimp \mi{Permitted}(a,\PlaySports)$, then in a world where 
$\mi{Student}(a)$ and $\mi{Female}(a)$ are true, then so is 
$\mi{Permitted}(a,\PlaySports)$.  Similarly, if we have a 
rule that says  $\mi{Faculty}(a) \land \mi{Female}(a) 
\rimp \neg \mi{Permitted}(a,\PlaySports)$, then in a world 
where $\mi{Faculty}(a)$ and $\mi{Female}(a)$ are true,  
$\neg \mi{Permitted}(a,\PlaySports)$ as well.  Of course, 
in a world $\mi{Faculty}(a)$, $\mi{Student}(a)$, and  
$\mi{Female}(a)$ are all true, both 
$\mi{Permitted}(a,\PlaySports)$ and  
$\neg \mi{Permitted}(a,\PlaySports)$ are true; this is an 
impossible world. 
This type of implicit description (and hence, impossible-worlds 
structure) should  
also be useful for characterizing large databases, when it is not 
possible to list all the tables explicitly.  
\end{example}

\section{Conclusion} 

Many solutions have been proposed to the logical omniscience problem, 
differing as to the intuitions underlying the lack of logical 
omniscience.  
There has been comparatively little work on comparing approaches. 
We have attempted to do so here, focussing on two aspects, 
expressiveness and pragmatics, for four popular approaches.  

In comparing the expressive power of the approaches, we started with the  
well-known observation that the approaches are equi-expressive in the 
propositional case.  
However, this observation is true only if we 
allow the agent not to consider any world possible.  
If we require that at least one world be possible, then we get a 
difference in expressive power.  
This is particularly relevant when we have probabilities, because 
there has to be at least one world over which to assign 
probability.  
Indeed, when considering logical omniscience in the presence of 
probability, there can be quite significant differences in expressive 
power between the  
approaches, particularly awareness and impossible worlds.  

Considering the pragmatics of logical omniscience, we identified  
some 
guiding principles for choosing an approach to model a situation, 
based on the source of the lack of logical omniscience in that 
situation.  
As we show, coming up with an appropriate structure can be nontrivial. 
We illustrate a general approach to deriving an impossible-worlds 
structure based 
on an implicit description of the situation, which  
seems to be appropriate for a number of situations of interest.  
Our discussion suggests that the impossible-worlds approach may be
particularly appropriate for representing an agent's subjective view of
the world.

\bibliographystyle{chicagor}
\bibliography{z,joe,riccardo2} 

\end{document}